\documentclass[11pt]{amsart}
\usepackage{amsthm,amssymb,amsmath,amscd,amsfonts,enumitem,graphicx,tikz,hyperref}
\hypersetup{final}
\usepackage[utf8]{inputenc}
\usepackage[english]{babel}
\usepackage[margin=1in]{geometry}

\theoremstyle{plain}

\theoremstyle{definition}

\makeatletter
\let\@@pmod\pmod
\DeclareRobustCommand{\pmod}{\@ifstar\@pmods\@@pmod}
\def\@pmods#1{\mkern4mu({\operator@font mod}\mkern 6mu#1)}
\makeatother

\newcommand{\mbbC}{\mathbb{C}}

\newcommand{\mbbF}{\mathbb{F}}

\newcommand{\mbbX}{\mathbb{X}}

\newcommand{\mbbZ}{\mathbb{Z}}

\newcommand{\symdiff}{\ominus}
\newcommand{\transpose}{T}

\newcommand{\xxz}{${}_X{\mbbX}_Z$}
\newcommand{\headerxxz}{\texorpdfstring{${}_X{\mbbX}_Z$}{X\_{\bf X}\_Z}}
\newcommand{\headertwoq}{\texorpdfstring{$2q$}{2q}}

\newcommand{\twovec}[2]{\left( \! \begin{array}{c} #1 \\ #2 \end{array} \! \right)}
\newcommand{\twomat}[4]{\left( \! \begin{array}{cc} #1 & #2 \\ #3 & #4 \end{array} \! \right)}

\begin{document}
\title{Generalized Haah Codes and Fracton Models}

\author{Kevin T. Tian}
\address{Department of Mathematics, University of California, Santa Barbara, CA 93106, USA}
\email{ktian@math.ucsb.edu}

\author{Zhenghan Wang}
\address{Microsoft Station Q and Department of Mathematics, University of California, Santa Barbara, CA 93106, USA}
\email{zhenghwa@microsoft.com;zhenghwa@math.ucsb.edu}

\begin{abstract}

Entanglement renormalization group flow of the Haah cubic code produces another fracton model with 4 qubits per lattice site,  dubbed as the Haah B-code.  We provide a schema that generalizes both models to stabilizer codes on any finite group  with $2q$ qubits per site and labeled by multi-subsets of the finite group and a set of pairwise-commuting binary matrices.

\end{abstract}


\maketitle

\section{Introduction}

Topological quantum computing with non-abelian anyons is susceptible to thermal anyon errors at finite temperature.  The search for robust quantum memory at finite temperature leads to the discovery of fracton models including the Haah (cubic) code \cite{haah2011local}.  In the study of  entanglement renormalization group (ERG) flow of the Haah code, a cousin of the Haah code is discovered, which is at least as interesting as the original Haah code \cite{haah2014bifurcation} \footnote{There are no string operators in the Haah B-model as well.  We thank Jeongwan for the following argument: if so, then his ERG equation $VH_A(a)V^{\dagger}\cong H_A(2a)+H_B(2a)$ would lead to string operators of Haah A-code, contradiction}.  Following \cite{haah2014bifurcation}, we will refer to the Haah cubic code as the Haah A-code, and the new model as the Haah B-code.  In this paper, we generalize the Haah codes in two directions: the Haah A-code and Haah-B code have two or four qubits on each site, respectively, first we generalize them to a code with $2q$ qubits on each site for any natural number $q$; secondly we generalize the Haah stabilizers to a variety of stabilizers labeled by multi-subsets of finite groups, thus to all three manifolds following \cite{tian2018haah}.

Large spin models such as spin-$\frac{3}{2}$ are potentially effective description of topological structures in certain materials.  Therefore, it is more than an academic curiosity to study many-body systems with more than one qubit per site.  The insight from Haah codes is that even with stabilizer codes, models with more than one qubit per site lead to intriguing new phenomena that potentially lie beyond quantum field theory description \cite{haah2014bifurcation}.  In this paper, we present a systematic construction for a large class of such models.

We consider lattices whose sites are labeled by finite groups $G$, hence each site is identified with a group element $g\in G$.  Fixed an integer $q\geq 1$, there live $2q$ qubits on each site $g$.  To define the codes, we choose two families $L=\{A_i\}, i=1,...,q, R=\{B_j\}, j=1,...,q$, of $q$ subsets of $G$, which play multiple roles and a $q\times q$ $\mbbF_2$ matrix $\chi$.
For Haah A-code, $G$ is the finite torus $\mbbZ_l\times \mbbZ_m\times \mbbZ_n$ with generators $\{x,y,z\}$ and unit $1$, and the Haah A-code is recovered from $A_1 = \{1,x,y,z\}$ and $B_1 = \{1, xy, xz, yz\}$ using the notation in \ref{subsec: xXz stabilizer}. The Haah B-code is recovered from $A_1 = \{x,z\}, A_2 = \{1,x\}$ and $B_1=\{x,y\}, B_2 = \{1,y\}$.

We do not touch on many interesting properties of our models in this paper such as degeneracy, excitations, and renormalization flow, which will be left to the future.

\section{Generalized Haah codes}

The following stabilizer construction plays a central role in generalizing Haah codes.

\subsection{\headerxxz  stabilizers}
\label{subsec: xXz stabilizer}

Fix a finite group $G$, an integer $q\geq 1$, and $q$ subsets $M=\{M_i\}, i=1,...,q$, of $G$.   Let $\mbbF_2$ be the field of two elements $\{0,1\}$.  We will identify any subset $M_i$ of $G$ as an element $M_i=\oplus_{m\in M_i} m$ of $\mbbF_2[G]$, the $\mbbF_2$-group algebra of $G$ \footnote{The $\mbbF_2$-group algebra $\mbbF_2[G]$ is the vector space consisting of all formal sums $X=\oplus_{g\in G}\epsilon_g g$ with the obvious product, where $\epsilon_g$=0 or 1. We identify the element $X=\oplus_{g\in G}\epsilon_g g$ with the subset of $G$ consisting of all elements with $\epsilon_g=1$.}. 
Conversely, an element of the $\mbbF_2$-group algebra induces a subset $S$ of $G$ corresponding of those elements $g$ for which the coefficient of $g$ is 1. We will use these views interchangeably.

Let 
\[ L(G,2q)=\otimes_{g\in G} (\mbbC^2)^{\otimes (2q)}\]
be the Hilbert space that associates $2q$ qubits to each group element $g$.
We will index the $2q$ qubits as $+1,...,+q,-1,..,-q$ and imagine the first $q$ qubits living above the \lq\lq lower" $q$ qubits in a bilayer mental picture.  Given a Pauli operator $P$, then $P(g,\pm i)$ denotes the Pauli operator $P$ acting on the qubit of $L(G,2q)$ indexed by $\pm i$ associated to the group element $g$, while $P(S,\pm i)=\prod_{g\in S}P(g, \pm i)$ for a subset $S$.

Suppose $A = (A_1, \ldots, A_q)$ and $B = (B_1, \ldots, B_q)$ are two vectors of $\mbbF_2^q$ (equivalently, two families of $q$ subsets of $G$), and $C$ is a $(q\times q)-\mbbF_2$ matrix. Then let $(CA)_i$ (resp. $(CB)_i$) denote the $i$-th component of the vector $CA$ (resp. $CB$), which is an element of $\mbbF[G]$ and thus defines a subset of $G$.

Then $\{A, B, C\}$ determine two \xxz stabilizers \footnote{We imagine that $X$ consists of two ribbons of width $q$, and the $Z$ stabilizer is defined on the main diagonal, while $X$ stabilizer on the minor diagonal.} on $L(G,2q)$ for each group element $g$ by

\begin{align*}\label{eq:XXZ}
  Z^C_g =& \prod_{k=1}^q Z_{(g (CA)_k, +k)} Z_{((C^\transpose B)_k g, -k)} \\
  X^C_g =& \prod_{k=1}^q X_{(\overline{(C^\transpose B)_k} g, +k)} X_{(g \overline{(CA)_k}, -k)}
\end{align*}

\subsection{Generalized Haah A-codes}

Generalized Haah A-codes are studied in \cite{tian2018haah}, where $q=1$, so each group element is associated with a bi-qubit $\mbbC^2\otimes \mbbC^2$. In this case, the \xxz codes reduce to the LR codes.  

In this section, we define a distance on a finite group and use it to discuss the issue of locality of the resulting Hamiltonians.  

\subsubsection{LR word metric on a group}

Fix a group $G$ and two inverse-closed subsets $L,R \subset G$, that is $L = \overline{L}$ and $R = \overline{R}$. We define a distance metric on $G$ as follow: 

Given $g,h \in G$, suppose there exist $s_1, \ldots, s_a \in L$ and $t_1, \ldots, t_b \in R $ such that $g = s_1 \cdots s_a h t_1 \cdots t_b$, then we define the distance $d(g,h)$ between $g$ and $h$ to be the minimum of $l = a+b$; otherwise $d(g,h)=\infty$. 
This is a modified word metric, and can be verified to be a distance via the same arguments as the word metric. 

\subsubsection{Locality of LR Hamiltonians}

If $G$ is non-abelian, then this metric is not necessarily left- or right-translation-invariant, and can also be very far from the natural metric on a Cayley graph. Under this metric, our LR Hamiltonians \cite{tian2018haah} are supported on balls of radius 1, hence local.

The metric depends on $L$ and $R$, but these ``shapes" are allowing an interaction between qubits that are related by a left-translation by $L$ or a right-translation by $R$.

\subsection{Generalized Haah B-codes}

\begin{figure}
  \tikzset{every picture/.append style={scale=0.85,font=\footnotesize}}
  \centering
  \begin{tikzpicture}
  \tikzset{vertexlabel/.style={text=black,fill opacity=1,fill=white,text opacity=1}}
  \draw[gray](2,0,0)--(0,0,0)--(0,2,0);
  \draw[gray](0,0,0)--(0,0,2);
  \draw[thick](2,2,0)--(0,2,0)--(0,2,2)--(2,2,2)--(2,2,0)--(2,0,0)--(2,0,2)--(0,0,2)--(0,2,2);
  \draw[thick](2,2,2)--(2,0,2);
  \node[vertexlabel] at (2,2,2) {$ZZZZ$};
  \node[vertexlabel] at (0,2,2) {$ZIII$};
  \node[vertexlabel] at (2,0,2) {$IIIZ$};
  \node[vertexlabel] at (2,2,0) {$IZZI$};
\end{tikzpicture}
\begin{tikzpicture}
  \tikzset{vertexlabel/.style={text=black,fill opacity=1,fill=white,text opacity=1}}
  \draw[gray](2,0,0)--(0,0,0)--(0,2,0);
  \draw[gray](0,0,0)--(0,0,2);
  \draw[thick](2,2,0)--(0,2,0)--(0,2,2)--(2,2,2)--(2,2,0)--(2,0,0)--(2,0,2)--(0,0,2)--(0,2,2);
  \draw[thick](2,2,2)--(2,0,2);
  \node[vertexlabel] at (2,2,2) {$IZIZ$};
  \node[vertexlabel] at (0,2,2) {$ZZII$};
  \node[vertexlabel] at (2,0,2) {$IIZI$};
  \node[vertexlabel] at (2,2,0) {$ZIZZ$};
\end{tikzpicture}

\begin{tikzpicture}
  \tikzset{vertexlabel/.style={text=black,fill opacity=1,fill=white,text opacity=1}}
  \draw[gray](2,0,0)--(0,0,0)--(0,2,0);
  \draw[gray](0,0,0)--(0,0,2);
  \node[vertexlabel] at (0,0,0) {$XXXX$};
  \draw[thick](2,2,0)--(0,2,0)--(0,2,2)--(2,2,2)--(2,2,0)--(2,0,0)--(2,0,2)--(0,0,2)--(0,2,2);
  \draw[thick](2,2,2)--(2,0,2);
  \node[vertexlabel] at (2,0,0) {$IIXI$};
  \node[vertexlabel] at (0,2,0) {$IXII$};
  \node[vertexlabel] at (0,0,2) {$XIIX$};
\end{tikzpicture}
\begin{tikzpicture}
  \tikzset{vertexlabel/.style={text=black,fill opacity=1,fill=white,text opacity=1}}
  \draw[gray](2,0,0)--(0,0,0)--(0,2,0);
  \draw[gray](0,0,0)--(0,0,2);
  \node[vertexlabel] at (0,0,0) {$IXIX$};
  \draw[thick](2,2,0)--(0,2,0)--(0,2,2)--(2,2,2)--(2,2,0)--(2,0,0)--(2,0,2)--(0,0,2)--(0,2,2);
  \draw[thick](2,2,2)--(2,0,2);
  \node[vertexlabel] at (2,0,0) {$IIXX$};
  \node[vertexlabel] at (0,2,0) {$XIII$};
  \node[vertexlabel] at (0,0,2) {$XXXI$};
\end{tikzpicture}
  \caption{Haah B-code stabilizers}
  \label{fig:haahB}
\end{figure}
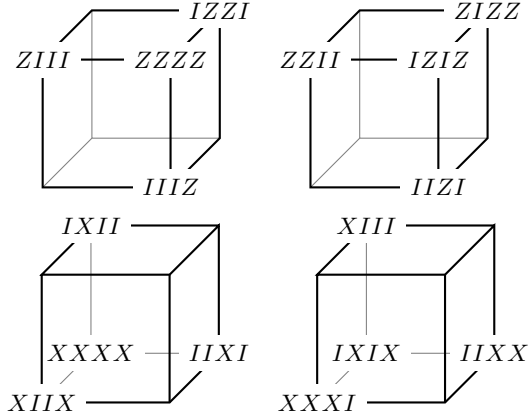

The Haah B-Code stabilizers are shown in Fig. \ref{fig:haahB} (copied from \cite{haah2014bifurcation}).  This is the case $q=2$, so each group element is associated with $4$-qubits.  One nice feature for Haah B-code is the self-replication in ERG flow \cite{haah2014bifurcation}:

$$VH_B(a)V^{\dagger}\cong 2H_B(2a),$$ which differs from the Haah A-code bifurcation 
$$UH_A(a)U^{\dagger}\cong H_A(2a)+H_B(2a).$$

\subsubsection{Generalizing LR codes  with \headertwoq qubits per site}\label{subsec::2qqubits}

The \xxz code generalizes Hastings' LR code \cite{tian2018haah} from 2 qubits per site to $2q$ qubits per site. We will label these qubits $+i$ and $-i$ for $i=1, \ldots, q$. Instead of 2 sets $S_1$ and $S_2$, we specify $2q$ sets, $A_1, \ldots, A_q$ and $B_1, \ldots, B_q$.
The stabilizers at a site $g$ are given by

\begin{align*}
  Z_g =& (Z_{(g A_1, +1)} \cdots Z_{(g A_q, +q)})(Z_{(B_1 g, -1)} \cdots Z_{(B_q g, -q)}) \\
  X_g =& (X_{\overline{B_1} g, +1} \cdots X_{(\overline{B_q}, +q)})(X_{(g\overline{A_1}, -1)} \cdots X_{g\overline{A_q}, -q)})
\end{align*}
where $Z_{(g,\pm i)}$ (resp. $X_{(g,\pm i)}$) denotes a Pauli Z (resp. Pauli X) acting on the qubit $\pm i$ at the site $g$.

For any fixed $i : \, 1 \leq i \leq q$, focusing on just qubits $+i$ and $-i$ gives us a 2 qubit LR code with sets $S_{i}$ and $S_{q+i}$. Thus, commutation of the stabilizers follows from the same argument as the 2-qubit LR code.

\subsubsection{\headerxxz code with \texorpdfstring{$q=2$}{q=2}: 4 qubits and 4 stabilizers per site}\label{subsec::fourstabilizer}

The above construction for $q=2$ takes $4$ sets $A_1, A_2, B_1$, and $B_2$, and produces two stabilizers for a code with 4 qubits per site. We now derive a second pair of stabilizers by defining 4 sets $A_1', A_2', B_1'$, and $B_2'$ based on $A_1, A_2, B_1$, and $B_2$. With the $A_i$'s and $B_i$'s, we create a second pair of stabilizers per site $g$, $Z'_g$ and $X'_g$ (denoting, respectively, the $Z$ and $X$ stabilizers from section \ref{subsec::2qqubits}, with the $A_i'$ and $B_i'$ sets below replacing the $A_i$ and $B_i$ sets). The construction of the $A_i'$ and $B_i'$ sets is given by: 
\[
    \begin{array}{rclcrcl}
        A_1' &=& A_1 \symdiff A_2 & \hspace{2em} & B_1' &=& B_1 \symdiff B_2 \\
        A_2' &=& A_1 & & B_2' &=& B_1
    \end{array}
\]
where $S \symdiff T$ indicates a symmetric difference
. Note that with the symmetric difference as addition, the system of subsets of $G$ is isomorphic to $\mbbF_2[G]$ as an $\mbbF_2$-module, with multiplication by 0 returning an empty set, and 1 acting by identity. In that case, the following equation also describes the relationship between the $S$ sets and the $T$ sets:
\begin{equation} \label{eq:st}
  \twovec{A_1'}{A_2'} = \twomat{1}{1}{1}{0} \twovec{A_1}{A_2}, \text{ and }
  \twovec{B_1'}{B_2'} = \twomat{1}{1}{1}{0} \twovec{B_1}{B_2}
\end{equation}
In particular, note that they are related by the same symmetric matrix.

Each $Z_g$ commutes with each $X_h$ for all $g,h$, and each $Z_g$ commutes with each $X_h$ for all $g,h$, since these are just copies of the two stabilizer LR code from section \ref{subsec::2qqubits}. The remaining commutation relation that needs to be checked is between the following 2 stabilizers:
\begin{align*}
    Z_g =& Z_{(gA_1, +1)} Z_{(gA_2, +2)} Z_{(B_1g, -1)} Z_{(B_2g, -2)} \\
    X'_h =& X_{(\overline{B_1'}h, +1)} X_{(\overline{B_2'}h, +2)} X_{(h\overline{A_1'}, -1)} X_{(h\overline{A_2'}, -2)}
\end{align*}

Fix $g,h \in G$ for the stabilizers above, and then fix $u,v \in G$ such that $gu = v^{-1}h$. Then these two stabilizers overlap on qubit $+i$ when $u \in A_i$ and $v \in B_i'$, and on qubit $-i$ when $v \in B_i$ and $u \in A_i'$ (note that $gu = v^{-1}h$ is equivalent to $v g = h u^{-1}$).

If we let $u_i$ indicate for $i = 1,2$ whether $u$ is in the set $A_i$, and let $v_j$ indicate for $j = 1,2$ whether $v$ is in the set $B_j$, then these indicators also determine the membership of $u$ and $v$ in the $A_i'$ and $B_i'$ sets respectively. Furthermore, we can write down an expression for each qubit based on the $u_i$ and $v_j$ for whether the two stabilizers overlap on that qubit. As an example, for qubit +1, this is simply $u_1 v_2$ ($u$ must be in $A_1$ and $v$ must be in $B_1' = B_2$), and for qubit +2, this is $u_2 v_1 + u_2 v_2$ ($u$ must be in $A_2$ and $v$ must be in $B_2' = B_1 \symdiff B_2$). The total number of overlaps (after summing the formulas for all 4 qubits) is then given by $2u_1 v_2 + 2 u_2 v_1 + 2 u_2 v_2$, which we see must always be even, and thus the two stabilizers $Z_g^A$ and $X_h^B$ commute.

\subsubsection{Recovering Haah B-Code}

The Haah B-Code fits in this 4 qubit, 4 stabilizer LR code model with the following sets:

\[
    \begin{array}{rclcrcl}
        A_1 &=& \{1,-y\} & \hspace{2em} & B_1 &=& \{1,-x\} \\
        A_2 &=& \{1,-x\} & \hspace{2em} & B_2 &=& \{1,-z\}
    \end{array}
\]

\subsection{\headerxxz code for \texorpdfstring{$2q$}{2q} qubits and \texorpdfstring{$2q$}{2q} stabilizers per site}\label{subsec::2qqubitsstabilizers}

The \xxz code with $2q$ qubits per site of $G$, on the Hilbert space $L(G,2q)$, is defined by 2 elements $A = (A_1, \ldots, A_q)$ and $B = (B_1, \ldots, B_q)$ of $\mbbF_2^q$, along with $q$ distinct pairwise-commuting $(q\times q)-\mbbF_2$-matrices $\{C_1, \ldots, C_q\}$ \footnote{We may also pick fewer than $q$ matrices, in which case we will get correspondingly fewer pairs of stabilizers.}.

Let us state how we specify each of $q$ pairs of stabilizers. Given a $(q \times q)-\mbbF_2$ matrix $\chi$, let $\chi^\transpose$ denote the tranpose. Then the stabilizers corresponding to the matrix $\chi$ are given by

\begin{align*}\label{eq:XXZ}
  Z^\chi_g =& \prod_{i=k}^q Z_{(g (\chi A)_k, +k)} Z_{((\chi^\transpose B)_k g, -k)} \\
  X^\chi_g =& \prod_{i=k}^q X_{(\overline{(\chi^\transpose B)_k} g, +k)} X_{(g \overline{(\chi A)_k}, -k)}
\end{align*}

These recover the stabilizers from section \ref{subsec::2qqubits} when we choose $\chi$ to be the identity matrix. With $q$ matrices $C_1, \ldots, C_q$, this yields $q$ pairs of stabilizers.

The $Z$ and $X$ stabilizers corresponding to the same matrix $C_i$ commute due to section \ref{subsec::2qqubits}, so it suffices to check that the $Z$ stabilizer and $X$ stabilizers from different matrices commute. 

Fix $g,h,u,v \in G$ satisfying that $gu=v^{-1}h$. We will show that the following 2 stabilizers commute:

\begin{align*}
  Z^{C_i}_g =& \prod_{k=1}^q Z_{(g (C_i A)_k, +k)} Z_{((C_i^\transpose B)_k g, -k)} \\
  X^{C_j}_g =& \prod_{k=1}^q X_{(\overline{(C_j^\transpose B)_k} g, +k)} X_{(g \overline{(C_j A)_k}, -k)}
\end{align*}

These two stabilizers overlap on the $+k$ qubit at $gu = v^{-1} h$ if $u \in (C_i A)_k$ and $v \in (C_j^\transpose B)_k$, and at the $-k$ qubit at $vg = h u^{-1}$ if $v \in (C_i^\transpose B)_k$ and $u \in (C_j A)_k$. Let $u_A \in \mbbF_2^q$ be a vector whose $k$-th entry is 1 if $u \in A_k$ and 0 otherwise. Similarly let $v_B \in \mbbF_2^q$ be the vector whose $k$-th entry is 1 if $v \in B_k$ and 0 otherwise. 

The membership of $u$ in $(C_i A)_k$ is determined by the $k$-th component of the vector $C_i u_A$, so we can write the following expression to indicate intersection on the $+k$ qubit: $(C_i u_A)_k (C_j^\transpose v_B)_k$. Then the number of intersections is given by
\begin{align*}
  \sum_k (C_i u_A)_k (C_j^\transpose v_B)_k + \sum_k (C_i^\transpose v_B)_k (C_j u_A)_k
    =& (C_j^\transpose v_B)^\transpose (C_i u_A) + (C_i^\transpose v_B)^\transpose (C_j u_A) \\
    =& v_B^\transpose C_j C_i u_A + v_B^\transpose C_i C_j u_A
\end{align*}
The two terms are equal since $C_j$ and $C_i$ commute, and so the stabilizers commute.

In addition, we note that if $G$ is an abelian group, then the $C_i$ may instead be matrices with entries from the group algebra $\mbbF_2[G]$ rather than just $\mbbF_2$, and commutativity of stabilizers follows from a similar calculation as the above.

\subsection{Qudit generalization}

Analogous to the qudit generalization of LR codes, the \xxz code has a generalizations to qudits $\mbbC^d$ as well. Let $U$ and $V$ be two operators acting on individual qudits, with eigenvalues $\{\omega_d^m\}, m = 0, 1, \ldots, d-1, \omega_d = e^{\frac{2\pi i}{d}}$, and $UV = \omega_d VU$.

We now draw our vectors $A,B$ from $\mbbZ_d^q$. Correspondingly, our subsets are now multisets in which each element's count only matters modulo $d$. Similarly, we take $q$ $(q \times q)$ $\mbbZ_d$ matrices. Let $m_A$ and $m_B$ be two maps from $G$ to $\{1, 2, \ldots, d-1\}$. Then for a matrix $\chi$, we then replace $Z_g$ and $X_g$ with

\begin{align*}
    U^{\chi}_g =& \prod_{k=1}^q \prod_{u \in (\chi A)_k} U_{(gu,+k)}^{m_A(u)} \prod_{v \in (\chi^\transpose B)_k} V_{(vg,-k)}^{m_B(v)} \\
    V^{\chi}_g =& \prod_{k=1}^q \prod_{v \in (\chi^\transpose B)} V_{(v^{-1} g, +k)}^{m_B(v)} \prod_{u \in (\chi A)_k}  V_{(g u^{-1}, -k)}^{m_A(u)}
\end{align*}

Commutativity of $U^{C_i}_g$ and $V^{C_j}_h$ follows by a similar computation as the qubit case, where the total phase from interchanging the + qubits equals the total phase from interchanging the - qubits. 

\section{Fracton Models}

Haah codes stirred up great interests in fractons and represent the most interesting fracton models when formulated as exactly solvable Hamiltonians. But in what sense that the Haah models represent topological phases is a very interesting open question.  From symmetry point-view, fracton models interpolate between global symmetries of co-dimension=$0$ and gauge symmetries of co-dimension=$n$ in space-time, possessing sub-dimensional symmetries.  

\subsection{Locality of the \headerxxz Hamiltonians}

To generalize to the $2q$ qubit case, we just have to take unions over all the left-multiplied sets and right-multiplied sets to form our left and right sets for this metric, and the stabilizers are still supported on balls of radius 1 (note that every element in a left-multiplied set came from some left-multiplied set, and similar for right-multiplied sets).

\subsection{Resonant scale and degeneracy}

One of the unusual features of the Haah code is its intriguing degeneracy formula. When the lattice size $L$ is a power of 2, the log-degeneracy is $4L-2$.  We speculate on a possible explanation of the salient degeneracy pattern in Haah codes using the notion of resonant scale.

Something somewhat similar in the LR code \cite{tian2018haah} happens when $G = \mathbb{Z}_n$, and $S_1 = \{0,a\}$ and $S_2 = \{0,b\}$. If $\text{gcd}(a,b) > 1$, then the degeneracy can depend on whether $\text{gcd}(a,b)$ divided $n$. For fixed $a,b$, the degeneracy of this code also depends on the lattice size $n$.
The massive scale-up of the degeneracy is not found here, so it begs the question whether there is some resonant scales in these codes so that when the size of the lattice is just right compared to these scales, the degeneracy is exponentially high.

Specifically, in the simple LR code example, $\text{gcd}(a,b)$ is the \lq\lq smallest" element that we can make, and generates the subgroup of all the locations to which a particle at 0 can be moved. In the Haah code, combining tetrahedra allows one to generate sets of 4 particles, but only at distances $2^k$ for some $k$.

\subsection{ERG and Limit}

Fix an integer $q\geq 1$ and an infinite group $\Gamma$.  Given any two families of finite subsets $\{A_i\}_{i=1}^q, \{B_j\}_{j=1}^q$ of $\Gamma$, and a $(q\times q)$-$\mbbF_2$-matrix $C$, we have a Hamiltonian on $L(G_N,2q)$ for all finite quotients $G_N={\Gamma}/{N}$, where $N$ is an FIN (a finite index normal subgroup) of $\Gamma$.  If $N\subset M$ is a pair of FINs, we are interested in defining an ERG map from $L(G_M, 2q)$ to $L(G_N, 2q)$.  

In \cite{haah2014bifurcation}, Haah studied $\Gamma=\mbbZ^3$ and $N= (2a)\mbbZ \times (2a)\mbbZ\times (2a)\mbbZ$ and $M=a\mbbZ \times a\mbbZ\times a\mbbZ$.

An interesting discovery of Haah are two ad hoc disentangling local unitaries denoted as $U$ and $V$, respectively, in \cite{haah2014bifurcation} such that
$$VH_B(a)V^{\dagger}\cong 2H_B(2a),$$ and 
$$UH_A(a)U^{\dagger}\cong H_A(2a)+H_B(2a),$$
where $a$ is the lattice size.  Naively, we would guess that the disentangling unitaries have certain universal properties so that they will also disentangle our generalizations.  Preliminary numerical simulations suggest that this is not correct.  So what makes the Haah models special among all the generalizations?

A deeper property of Haah models is the increase of effective degrees of freedom under ERG, which does not occur for any renormalizable Lorentz-invariant quantum field theory.  Hence it is very interesting to study if there are some kinds of continuous limits of the generalized Haah models. 

\section*{Acknowledgments}

K.T. and Z.W. are partially supported by NSF grant  FRG-1664351. Both authors thank J. Haah for insightful comments.

\bibliography{haahB}

\begin{thebibliography}{1}

\bibitem{haah2011local}
Jeongwan Haah.
\newblock Local stabilizer codes in three dimensions without string logical
  operators.
\newblock {\em Physical Review A}, 83(4):042330, 2011.

\bibitem{haah2014bifurcation}
Jeongwan Haah.
\newblock Bifurcation in entanglement renormalization group flow of a gapped
  spin model.
\newblock {\em Physical Review B}, 89(7):075119, 2014.

\bibitem{tian2018haah}
Kevin~T Tian, Eric Samperton, and Zhenghan Wang.
\newblock Haah codes on general three manifolds.
\newblock {\em arXiv preprint arXiv:1812.02101}, 2018.

\end{thebibliography}
\bibliographystyle{plain}

\end{document}